%% file: main.tex
\documentclass[prl,twocolumn,aps,10pt]{revtex4-2}
\input{common.tex}

\begin{document}
\title{Information is localized in growing network models}
\input{frontmatter.tex}

\begin{abstract}
    % 5% of length, < 500 words
    Mechanistic network models can capture salient characteristics of empirical networks using a small set of domain-specific, interpretable mechanisms. Yet inference remains challenging because the likelihood is often intractable. We show that, for a broad class of growing network models, information about model parameters is localized in the network, i.e., the likelihood can be expressed in terms of small subgraphs. We take a Bayesian perspective to inference and develop neural density estimators (NDEs) to approximate the posterior distribution of model parameters using graph neural networks (GNNs) with limited receptive size, i.e., the GNN can only ``see'' small subgraphs. We characterize nine growing network models in terms of their localization and demonstrate that localization predictions agree with NDEs on simulated data. Even for non-localized models, NDEs can infer high-fidelity posteriors matching model-specific inference methods at a fraction of the cost. Our findings establish information localization as a fundamental property of network growth, theoretically justifying the analysis of local subgraphs embedded in larger, unobserved networks and the use of GNNs with limited receptive field for likelihood-free inference.
\end{abstract}
\maketitle

Mechanistic models with simple, domain-specific, interpretable rules can yield complex networks that capture salient characteristics of real-world data, such as heavy-tailed degree distributions~\cite{Vazquez2003a}, the small-world effect~\cite{Watts1998}, or properties of sexual contact networks~\cite{Hoffmann2025longitudinal}. However, fitting these models to data is challenging because the sequence in which different mechanisms were applied is typically unknown, rendering the likelihood intractable. Numerous inference approaches have been developed, but each has shortcomings. Approximate Bayesian computation draws posterior samples by comparing summary statistics~\cite{Raynal2023}, but it is computationally demanding and requires that we identify summary statistics of the network that are informative of the model parameters~\cite{Hoffmann2022}. The likelihood becomes tractable if we assume the order in which edges are added to the network is known~\citep{Arnold2021,Overgoor2019}. But this requires longitudinal data which is rarely available---we usually observe a snapshot in time rather than the entire history. Such rich data also pose privacy concerns~\citep{Musciotto2016}. Inferring the node order~\citep{Wiuf2006,Larson2023ReversibleNetwork,Zukowski2022NetworkHistory} typically expands the parameter space by orders of magnitude and is particularly challenging due to its combinatorial nature.

To address this challenge, we first show that information is localized for a broad class of growing network models, i.e., the likelihood, while still intractable, can be expressed in terms of small subgraphs, allowing parameter inference without observing the global network. Second, we leverage this theoretical result to design neural posterior density estimators (NDEs) for network data based on graph neural networks (GNNs). Finally, we demonstrate that inferences for simulated networks agree with theoretical predictions of information localization.

We focus on a sequential generative process where we grow an undirected graph $G$ with $n$ nodes comprising a set of edges $\edges$ by repeatedly applying the same rule until a termination criterion is satisfied. A rule is not limited to a single mechanism, e.g., network growth may comprise a mixture of mechanisms~\citep{wang2025approximatebayesianinferencemechanisms}. At each step $t$, we sample a source node $\order_t$ and add edges $\epsilon_{t}$ that connect it to the graph $\graph_{t-1}$ at the previous step. The growth is fully specified by
\begin{equation}
    \proba{\epsilon_{t}\mid \graph_{t - 1},\theta,\order_t}\proba{\order_t\mid\theta},
    \label{eq:conditional-full}
\end{equation}
where $\theta$ are model parameters. The first term encodes the mechanism for adding edges, and the second captures the source selection. Rules are often applied sequentially to newly added nodes such that $\order_t=t$, but nodes can be revisited in principle. We assume that edges are never deleted, and we call the growth \emph{monotonic}.

While the emergent global properties of mechanistic models are complex, most real-world networks arise from local processes~\cite{Vazquez2003a}. In practice, new edges do not depend on the entire graph---imagine having to consider billions of people to choose your friends. We consider a restricted, localized model. At each step $t$, we sample a source $\order_t$ and seed nodes $\seeds_{t}$ independent of $\graph_{t-1}$. We select targets for $\order_t$ by exploring the neighborhoods of seeds. We say that a rule is $k$-localized if new edges only depend on the subgraphs $\subgraph_{\seeds_{t}}^{(k)}\parenth{G_{t-1}}$ induced by the $k$-hop neighborhoods of seeds in $\graph_{t-1}$. The first term in \cref{eq:conditional-full} thus reduces to
\[
    \proba{\epsilon_{t}\mid \subgraph_{\seeds_{t}}^{(k)}\parenth{G_{t-1}},\order_t,\theta}.
\]
This class includes models for a wide range of networks, including evolution for protein interaction networks~\cite{Sole2002,Vazquez2003}, the web~\cite{Krapivsky2001}, social~\cite{Jackson2007}, small-world~\cite{Newman1999}, and citation networks~\cite{Golosovsky2019}.

\begin{figure}
    \centering
    \includegraphics{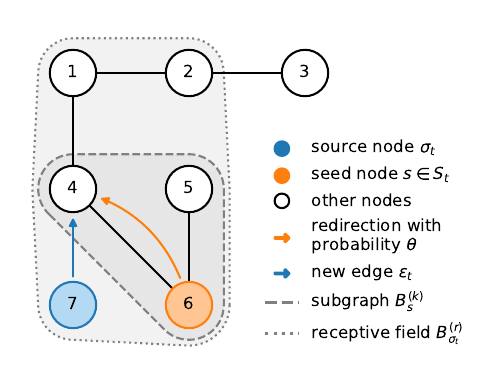}
    \caption{\emph{Illustration of network growth for a probabilistic redirection model~\cite{Krapivsky2001}.} At each step $t$, a new node $\order_t$ is added, and a single seed node $s$ is sampled uniformly at random. The new node is connected to $s$ with probability $1-\theta$ and to one of its uniformly chosen neighbors otherwise. New connections only depend on the direct neighborhood $B_{s}^{(1)}$ of the seed $s$, and the model is 1-localized. The receptive field of $\order_t$ is its three-hop subgraph $\subgraph^{(3)}_{\order_t}$. Nodes outside the receptive field cannot affect the new node and vice-versa.}
    \label{fig:redirection-illustration}
    % word count: [(150 / aspect ratio (1.32)) + 20 words] = 134
\end{figure}

If the sequence of source and seed nodes is known, the likelihood can be expressed as a product of local factors
\begin{equation}
    \proba{\edges\mid\theta,\order,S}=\prod_{t=1}^n \proba{\epsilon_{t}\mid \subgraph_{\seeds_{t}}^{(k)}\parenth{G_{t-1}},\order_t,\theta},
    \label{eq:factorized-likelihood}
\end{equation}
where $\order=\braces{\order_1,\ldots,\order_n}$ and $S=\braces{S_1,\ldots,S_n}$ are all source nodes and their corresponding seeds, respectively. The set of all edges is $\edges=\bigcup_{t=1}^n\epsilon_t$. Yet this longitudinal information is rarely available, and we need to consider all possible sequences $\order$ and seeds $S$ to obtain the marginal likelihood
\begin{equation}
    \proba{\edges\mid \theta}=\sum_{\order,S} \proba{\edges\mid\theta,\order,S}\proba{\order,S\mid\theta},
    \label{eq:marginal-likelihood}
\end{equation}
where the sum is over all possible permutations of $\order$ and seeds $S$. This task is not computationally feasible except for very small networks, and approximations include pseudo-marginal likelihood~\citep{Cantwell2021} and expectation-maximization methods~\citep{Larson2023ReversibleNetwork}. Nevertheless, the forms of \cref{eq:factorized-likelihood,eq:marginal-likelihood} are instructive. Because each factor only depends on subgraphs, both the conditional and intractable marginal likelihood can in principle be expressed in terms of local graph features.

How local are these features? Node $\order_t$ can only contribute to the likelihood in two ways. First, when it serves as a source for edges $\epsilon_{t}$ created at step $t$. Second, at a step $t'$ if it belongs to one of the seed subgraphs $\subgraph_{\seeds_{t'}}^{(k)}\parenth{\graph_{t'-1}}$ for a different node $\order_{t'}$. Any node that can affect or can be affected by $\order_t$ is thus at most a distance $2k+1$ away because the diameter of a $k$-hop subgraph is at most $2k$ and a further edge connects the subgraphs to their corresponding source node $\order_t$. The likelihood can therefore be expressed in terms of features extracted from $2k+1$-hop subgraphs, and we call the subgraph $\subgraph_{\order_t}^{(2k+1)}\parenth{\graph}$ the \emph{receptive field} of node $\order_t$.

For example, the probabilistic redirection model of \citet{Krapivsky2001} illustrated in \cref{fig:redirection-illustration} adds a new node $\order_t$ at each step and connects it to a random seed $s$ with probability $1-\theta$. Otherwise, $\order_t$ connects to a neighbor of $s$. Node 2 in \cref{fig:redirection-illustration} is in the receptive field of node 7 for the 1-localized redirection model despite being three hops removed. The connection between 4 and 7 could have arisen due to redirection from 1 to 4, and the probability of redirection to 4 depends on the neighbors of 1, which includes 2. However, there is no mechanism for node 3 to affect creation of the edge. Likewise, node 7 cannot affect edges of 3.

Importantly, because we consider monotonic growth models, any subgraph at step $t'$ is a superset of the subgraph centered on the same node at time $t<t'$. In other words, any node or edge that could have interacted with a node at any step will be part of its receptive field in the final graph $\graph$.

\begin{table}
    \centering
    \begin{tabular}{lll}
    \toprule
    \textbf{Model} & $\bm k$ & $\bm p$ \\
    \midrule
        Redirection~\cite{Krapivsky2001} & 1 & 1 \\
        Duplication \& mutation~\cite{Sole2002} & 1 & 2 \\
        Probabilistic copying~\cite{Lambiotte2016} & 1 & 1 \\
        Random connections~\cite{Callaway2001} & 0 & 1 \\
        Connected small world~\cite{Newman1999} & 0 & 1 \\
        Growing Tree~\cite{Cantwell2021} & -- & 1 \\
        Duplication \& complementation~\cite{Vazquez2003} & -- & 2 \\
        Jackson-Rogers~\cite{Jackson2007} & -- & 2 \\
        Watts-Strogatz~\cite{Watts1998} & -- & 1 \\
    \bottomrule
    \end{tabular}
    \caption{\emph{Information about parameters is localized in many mechanistic network models.} For each model, $k$ denotes the localization of the generative model, and $p$ is the number of parameters of the model to be inferred.}
    \label{tbl:models}

    % word count: 13 + 6.5 * number of lines (10) = 78.
\end{table}

As the likelihood remains intractable, we resort to simulation-based inference using neural density estimators (NDEs) \cite{Papamakarios2016epsilon-free}. For each mechanistic network model, we train an NDE $f\parenth{\theta\mid \graph}$ to approximate the posterior distribution $\proba{\theta\mid G}$ by minimizing the negative log probability loss
\begin{equation}
\mathcal{L}=-\E{\log f\parenth{\theta\mid \graph}},
\label{eq:neg-prob-loss}
\end{equation}
where $\E{\cdot}$ denotes the expectation under the prior predictive distribution~\citep{Papamakarios2016epsilon-free}. The NDE comprises three parts, and we use information localization to design the neural architecture.

First, we use graph isomorphism networks (GINs) with weighted shortcut connections as a local feature extractor. GINs are neural networks that learn node representations $\mathbf{H}^{(l)}$ by aggregating and transforming information captured by node representations $\mathbf{H}^{(l-1)}$ in the previous layer $l-1$~\cite{Xu2019a}. We use shortcut connections to stabilize gradients and facilitate training of the neural network~\citep{He2016resnet}. Each layer can be expressed as
\begin{equation}
\mathbf{H}^{(l)}=\gamma^{(l)} \mathbf{H}^{(l - 1)} + \phi^{(l)}\parenth{\tilde{\mathbf{A}} \mathbf{H}^{(l-1)}},
\label{eq:gin-convolution}
\end{equation}
where $\tilde{\mathbf{A}}$ is the adjacency matrix with added self-loops, $\gamma^{(l)}$ is a scale factor for shortcut connections in layer $l$, and $\phi^{(l)}$ is a multi-layer perceptron (MLP) applied row-wise after the $l^\text{th}$ aggregation. Self-loops are added to ensure the node's own features are included in the aggregation. Nodes often have features, such as demographics in social networks or atomic element type (e.g., H, C, O) in molecules. Because we only have structural information, we set the initial node representation $\mathbf{H}^{(0)}$ to the $n$-vector of ones. We obtain a graph-level representation $\boldsymbol\xi^{(0)}$ by averaging node features in the final layer, i.e.,
\begin{equation}
\boldsymbol\xi^{(0)}=n^{-1}\sum_{i=1}^n \mathbf{H}_i^{(\ell)}.
\label{eq:graph-representation}
\end{equation}

Second, we apply $M$ dense residual layers with the same functional form as the GIN layer MLPs to further transform the graph-level features $\boldsymbol{\xi}^{(0)}$ to $\boldsymbol{\xi}^{(M)}$. These layers employ the same MLP architecture as the GIN layers. They increase the expressivity of the NDE, allowing it to learn complex non-linear mappings between the aggregated graph statistics and the model parameters. Finally, we use a linear layer with softplus activation function $\mathrm{softplus}\parenth{x}=\log\parenth{1+\exp x}$ to estimate the concentration parameters of a beta distribution that approximates the posterior for each model parameter. This parametric approximation assumes that the posterior factorizes similar to a variational mean-field approximation. We employ beta distributions because all model parameters are probabilities. We used dense two-layer neural networks with eight hidden units and $\tanh$ activations for the MLP (wider or deeper architectures did not improve performance). See \suppref{appendix A}{section A of the online supplement} for details of the neural network architecture.

We trained NDEs with up to $L=5$ GIN layers in two steps. First, we sampled 10{,}000 graphs with 1{,}000 nodes each (a typical size for many network inference studies~\citep{Cantwell2021,Larson2023ReversibleNetwork,Raynal2023,Arnold2021}) from the prior predictive distribution as training data. Second, we minimized \cref{eq:neg-prob-loss} by averaging over mini-batches of 32 graphs using the Adam optimizer with an initial learning rate of $10^{-2}$~\citep{kingma2017adammethodstochasticoptimization}. The learning rate was halved if the loss on a separate validation set of 1{,}000 graphs did not decrease for 10~consecutive epochs, and training was terminated if the validation loss did not decrease for 25~epochs.
We repeated training five times for each configuration with different seeds to account for variability due to initialization, and we report results for the best training run based on the validation set. For fair comparison across varying depths $\ell$, we used $M=L-\ell$ dense layers after mean-pooling such that the number of model parameters is constant across architectures.

\begin{figure}
    \centering
    \includegraphics{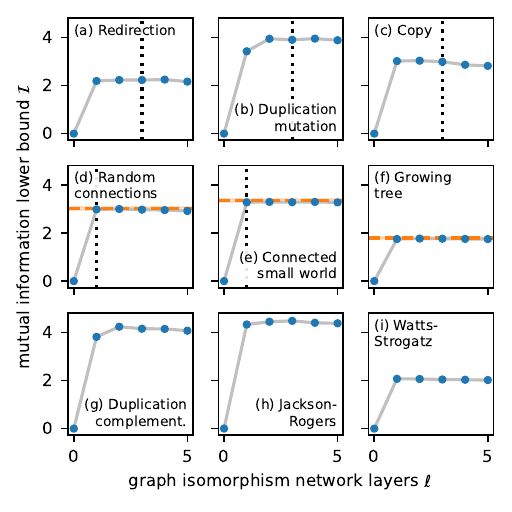}
    \caption{\emph{Neural density estimators can learn high fidelity posteriors.} Each panel shows an estimate of the variational mutual information $\mathcal{I}$ as a function of GIN depth $\ell$, based on a test set of 1{,}000 graphs for each of the models in \cref{tbl:models}. The larger $\mathcal{I}$, the better the inference. Five of the nine models are localized, and dotted vertical lines indicate the expected receptive field by which we expect $\mathcal{I}$ to saturate. Dashed horizontal lines correspond to posterior inference for models with tractable or approximated likelihood. Error bars are bootstrapped 95\% confidence intervals and are smaller than markers.}
    \label{fig:mi}
    % word count: [(150 / aspect ratio (1.0)) + 20 words] = 170
\end{figure}

We evaluated NDEs by estimating a variational lower bound $\mathcal{I}=\mathcal{H}\bracket{\pi}-\mathcal{L}$ for the mutual information between parameters and graph observations, where $\mathcal{H}\bracket{\pi}$ is the entropy of the prior $\pi$ and $\mathcal{L}$ is the negative log-probability loss in \cref{eq:neg-prob-loss}. The larger $\mathcal{I}$, the better the NDE is able to extract information and constrain parameters.

As shown in panel (a) of \cref{fig:mi} for the redirection model, the results improve with increasing GIN depth $\ell$ because the NDE is able to extract information from larger neighborhoods. Zero-depth GINs are unable to extract any information, and $\mathcal{I}=0$. The redirection model is 1-localized, and, as expected, the ability of NDEs to constrain parameters saturates as the depth $\ell$ approaches the size of the receptive field $2k+1$. We consider eight additional mechanistic network models listed in \cref{tbl:models} with varying degrees of localization (see \suppref{appendix B}{section B of the online supplement} for details). We use $\theta$ to denote parameters for all models, but they have distinct meaning for each.

\citet{Sole2002} introduced a duplication-mutation model for gene evolution where edges represent physical interactions between expressed proteins. At each step, a random gene $s$ is duplicated, and duplicated edges are retained with probability $\theta_1$. The new node $\order_t$ is dropped if isolated. Then, edges to other random nodes are added with probability $\theta_2 / n$, representing random mutations. A related copying model connects new nodes to a random seed $s$ and also its neighbors with probability $\theta$, but it does not add random long-range connections~\citep{Lambiotte2016}. The performance of NDEs for the two models is shown in panels~(b) and~(c). Both models are 1-localized because connections depend on direct neighbors of $s$ (random mutations do not depend on graph structure), and $\mathcal{I}$ saturates at $\ell=3$ as expected.

\citet{Callaway2001} proposed a mechanism that adds a node at every step and adds a random edge with probability $\theta$. The connected small world model of \citet{Newman1999} starts with a ring of nodes connected to nearest neighbors, and random connections are added with probability $\theta$. Both models are 0-localized because edges do not depend on the graph structure, and the posterior is tractable because the number of added edges follows a binomial distribution. As shown in panels~(d) and~(e), $\mathcal{I}$ saturates at $\ell=1$ and NDEs match the performance of analytic posteriors for these simple models. However, we observe a slight degradation for deep GINs with $\ell\geq4$.

\citet{Cantwell2021} considered a model of growing trees where new nodes make a single connection to existing nodes with probability proportional to $d^\theta$, where $d$ is the degree. The likelihood is not tractable because the node order $\order$ is unknown, and they inferred $\parenth{\theta,\order}$ jointly. However, history inference degrades for large $\theta$~\citep{Hoffmann2022} and is computationally expensive, taking 40~minutes for the test set. The model is not localized because selection of the seed depends on the degree of all nodes. Nevertheless, as shown in panel~(f), NDEs yield similar performance even at shallow depths. They also provide amortized inference, i.e., the model only needs to be trained once, taking 20~minutes for $\ell=2$. Evaluation on the test set takes less than a second.

The duplication-complementation~\citep{Vazquez2003}, Jackson-Rogers~\citep{Jackson2007}, and Watts-Strogatz~\citep{Watts1998} models are non-localized variants of the duplication mutation, copying, and connected small world models, respectively. They either remove edges or may, by chance, not make any connections to the neighborhood of seeds. Nevertheless, NDEs can learn parameters of these models, and performance quickly saturates with increasing depth $\ell$.

Five of the nine models are localized, and the ability of NDEs to constrain parameters saturates as the depth $\ell$ approaches the size of the receptive field $2k+1$.
This observation lends empirical support to our argument that local information is sufficient to constrain the parameters of local network growth. However, this is neither a necessary nor sufficient condition. For example, the NDE for the random connection model at depth $\ell=5$ performs slightly worse than the tractable posterior even though the latter can be expressed in terms of the node degree that depends only on the direct neighborhood. The NDE can approximate the posterior well in principle, and the deficiency is due to a lack of convergence during optimization (see \suppref{appendix C}{section C of the online supplement} for details). Further, each GIN layer aggregates node features from its direct neighborhood and cannot ``see'' important higher-order network structure, such as triangles. While GINs are provably the most expressive message-passing graph neural networks~\citep{Xu2019a}, a lack of further improvement with increasing depth could be due to insufficient expressive power of the neural networks. Higher-order graph neural networks can capture such structures but at increased computational cost~\citep{Morris2019}.

We have shown that information about the parameters of growing network models is often localized, i.e., local features are sufficient to infer parameters. Theoretical predictions of the receptive field of feature extractors required for inference agree with extensive simulations based on neural posterior estimates. For simple models with tractable likelihood, NDEs match the performance of exact inference. Inference could be further improved by considering more sophisticated density estimators than the parametric mean-field approximation we employed here, such as normalizing flows~\cite{Papamakarios2021}. The graph representation $\boldsymbol{\xi}^{(0)}$ in \cref{eq:graph-representation} is a mean of node representations, suggesting that principled parameter inference is feasible even when we only have access to samples of subgraphs. However, the receptive fields of nodes at the periphery of such samples are truncated, and future work should investigate the extent to which this affects parameter inference. NDEs are able to approximate the posterior of general network models even when information is not necessarily localized. They are a versatile tool for network inference that can be readily applied without having to develop model-specific methods. Once trained, NDEs provide inference in under a second compared to tens of minutes for model-specific methods. This paves the way for the development of powerful, off-the-shelf inference engines that make rigorous mechanistic network analysis accessible to the broader scientific community.

\paragraph{Acknowledgments} This work was supported by the National Institutes of Health under grant R01AI138901.

\bibliographystyle{apsrev4-2}
\bibliography{main.bib}

\ifsupplement
\else
\appendix
\input{supplement-content.tex}
\fi

\end{document}

%% file: common.tex
\usepackage[utf8]{inputenc}
\usepackage{verbatim}
\usepackage{bm}
\usepackage[colorlinks=true,allcolors=blue]{hyperref}
\usepackage{graphicx}
\usepackage{amsmath}
\usepackage{amssymb}
\usepackage{booktabs}
\usepackage{cleveref}
\usepackage{xcolor}

\newcommand{\parenth}[1]{\left(#1\right)}
\newcommand{\bracket}[1]{\left[#1\right]}
\newcommand{\braces}[1]{\left\{#1\right\}}
\newcommand{\proba}[1]{p\parenth{#1}}

\newcommand{\edges}{E}

\newcommand{\graph}{G}
\newcommand{\seeds}{S}
\newcommand{\subgraph}{B}
\newcommand{\order}{\sigma}
\newcommand{\E}[1]{\left\langle#1\right\rangle}
\newcommand{\abs}[1]{\left\vert#1\right\vert}
\newcommand{\dist}{\sim}

\newif\ifsupplement
\ifsupplement
    \newcommand{\suppref}[2]{#2}
\else
    \newcommand{\suppref}[2]{#1}
\fi

%% file: frontmatter.tex
\author{Till Hoffmann}
\email{thoffmann@hsph.harvard.edu}
\author{Jukka-Pekka Onnela}
\email{onnela@hsph.harvard.edu}
\affiliation{Department of Biostatistics, Harvard T.H.\ Chan School of Public Health, Boston, Massachusetts 02115, USA}

%% file: supplement-content.tex
\section{A.\ Neural architecture}

The neural density estimators (NDEs) comprise three parts: a graph isomorphism network (GIN) to extract local graph features with mean-pooling across all nodes to obtain graph-level features, dense layers to transform these features, and a density estimation head to approximate the posterior given a graph as input. We consider each in turn.

First, we learn node-level features using graph isomorphism network (GIN) layers defined in \suppref{\cref{eq:gin-convolution}}{eq.~(5) of the main text}. The multi-layer perceptron (MLP) is
\begin{align}
\phi^{(l)}\parenth{\mathbf{X}} = \bracket{\mathbf{b}_2^{(l)}}^\intercal
+ \tanh\parenth{\bracket{\mathbf{b}_1^{(l)}}^\intercal + \mathbf{X}\mathbf{W}_1^{(l)}}\mathbf{W}_2^{(l)},\label{eq:mlp}
\end{align}
where $\mathbf{b}$ are biases, $\mathbf{W}$ are weights, and $^\intercal$ denotes the transpose. After applying the GIN $\ell$ times, we mean-pool node features to obtain the graph-level representation $\boldsymbol\xi$ in \suppref{\cref{eq:graph-representation}}{eq.~(6) of the main text}.

Second, we transform $\boldsymbol\xi$ using $M$ scaled dense residual layers with the MLP in \cref{eq:mlp}.

Finally, we use independent parameterized beta distributions as a density estimation head such that
\[
    f\parenth{\theta\mid G} = \prod_{i=1}^p \frac{\Gamma\parenth{\alpha_i+\beta_i}}{\Gamma\parenth{\alpha_i}\Gamma\parenth{\beta_i}}\theta_i^{\alpha_i-1}\parenth{1-\theta_i}^{\beta_i-1},
\]
where $\Gamma$ denotes the gamma function. The non-negative concentration parameters are obtained using a linear layer with softplus transformation applied to the final graph-level representation, i.e.,
\[
\parenth{\alpha_i,\beta_i}^\intercal = \text{softplus}\parenth{\mathbf{b}_3 + \mathbf{W}_{3}\boldsymbol{\xi}^{(L)}},
\]
where $\mathrm{softplus}\parenth{x}=\log\parenth{1+\exp x}$.

\section{B.\ Growing network models\label{app:network-models}}

\newcounter{modelnum}
\newcommand{\modelnumref}{%
  \texorpdfstring{\addtocounter{modelnum}{1}\arabic{modelnum}}{\#}%
}

\subsection{\modelnumref.\ Probabilistic redirection}

\citet{Krapivsky2001} developed a probabilistic redirection graph to explain heavy-tailed degree distributions without global knowledge of the graph. At each step, a seed $s\in\graph_{t-1}$ is selected uniformly at random. With probability $1-\theta$, $\order_t$ connects to $s$. Otherwise, $\order_t$ connects to a randomly chosen neighbor of $s$. The model is one-localized because each step only depends on the direct neighborhood of $s$. When $\theta=0$, the model is equivalent to random attachment. When $\theta=1$, the model is equivalent to preferential attachment~\cite{Barabasi1999}. We use a $\text{Beta}\parenth{2,1}$ prior for the redirection parameter $\theta$ to reduce the likelihood of superhubs, i.e., nodes that attract the vast majority of edges.

\subsection{\modelnumref.\ Duplication divergence model with mutation}

\citet{Sole2002} introduced a model for gene evolution with two parameters and two stages at each step $t$. The first stage models duplication with divergence. A node $s\in \graph_{t-1}$ is randomly chosen as a seed. The new node $\order_t$ is connected to each neighbor of $s$ with probability $\theta_1$, and it is dropped if it is isolated. The second stage models mutations, and we sample the number of additional neighbors $z$ from a binomial distribution with $\abs{\graph_{t-1}}$ trials and success probability $\theta_2 / \abs{\graph_{t-1}}$, where $\abs{\graph_{t-1}}$ denotes the number of nodes in $\graph_{t-1}$. The new node $\order_t$ is then connected to $z$ randomly chosen nodes. The model is one-localized because connections only depend on the direct neighborhood of the seed $s$. We use weakly informative $\text{Beta}\parenth{2, 2}$ priors for both the connection and mutation parameters.

\subsection{\modelnumref.\ Probabilistic copying}

\citet{Lambiotte2016} introduced a model based on probabilistic copying that lends itself to analytic study. At each step, a seed $s$ is chosen uniformly at random and a connection between $\order_t$ and $s$ is created. Independently, $\order_t$ connects to each neighbor $u$ of $s$ with probability $\theta$. The model is one-localized, and we use a $\text{Beta}\parenth{2,2}$ prior for $\theta$.

\subsection{\modelnumref.\ Random connection model}

\citet{Callaway2001} introduced a simple yet rich network model. At each step, we first add a new node to the graph. Second, we pick any two nodes uniformly at random and connect them with probability $\theta$. The model is zero-localized because it does not depend on the graph structure. We use a $\text{Beta}\parenth{1,1}$ prior for the connection probability $\theta$ and initialize the graph with a single node. Due to its simplicity, the posterior can be obtained in closed form. The likelihood is a sequence of $n-1$ Bernoulli trials with success probability $\theta$, and the posterior is
\[
    % negative trials: n - 1 - |E|, positive trials |E|, add one for the Beta(1, 1) prior.
    \theta\mid\graph\dist\text{Beta}\parenth{1 + \abs{\edges}, n - \abs{\edges}},
\]
where $\abs{\edges}$ is the number of edges.

\subsection{\modelnumref.\ Connected small world}

\citet{Newman1999} introduced a model for graphs with short diameter and high clustering coefficients that is easy to study analytically. The initial graph is a ring of $n$ nodes, and each node is connected to its $z$ nearest neighbors. For each deterministic edge in the ring, we add a random edge with probability $\theta$. The model is zero-localized because it does not depend on the graph structure. We use a $\text{Beta}\parenth{1,1}$ distribution for the connection probability and set $z=4$. The number of random edges is $\abs{\edges} - \frac{n z}{2}=\frac{n\parenth{\bar d - z}}{2}$, where $\bar d=\frac{2\abs{\edges}}{n}$ is the mean degree. The model has a tractable likelihood because the number of random edges follows a binomial distribution $\text{Binomial}\parenth{\frac{n z}{2},\theta}$, and the posterior is
\begin{align}
\theta\mid\graph&\dist\text{Beta}\parenth{1 + \frac{n\parenth{\bar d - z}}{2}, 1 + \frac{nz}{2} - \frac{n\parenth{\bar d - z}}{2}}\nonumber\\
    &\dist\text{Beta}\parenth{1 + \frac{n\parenth{\bar d - z}}{2}, 1 + \frac{n\parenth{2z-\bar d}}{2}}.
    \label{eq:connected-small-world-posterior}
\end{align}

\subsection{\modelnumref.\ Growing tree model}

\citet{Cantwell2021} considered a model of growing trees. At each step, a new node is added to the graph, and it is connected to an existing node $s$ with probability proportional to $d_s^\theta$, where $d_s$ is the number of existing connections of $s$ and $\theta$ is an exponent controlling the degree of preferential attachment. The model is not localized because the seed node $s$ depends on the graph structure. For consistency with other parameters, we use a $\text{Beta}\parenth{1, 1}$ prior although the exponent can be any real number in principle.

\subsection{\modelnumref.\ Duplication divergence with complementation}

\citet{Vazquez2003} proposed an alternative model for gene evolution. Similar to the duplication divergence model with mutation, it has two parameters and two stages at each step. First, a node $s\in\graph_{t-1}$ is chosen. For each neighbor $u$ of $s$, we sample an indicator $z\dist\text{Bernoulli}\parenth{0.5}$. If $z=1$, we connect $\order_t$ to $u$ with probability $\theta_1$. Otherwise, we remove the connection between $s$ and $u$ with probability $1-\theta_1$. Second, we connect $s$ and $\order_t$ with probability $\theta_2$. The model is not localized because edges can be deleted, but its structure is very similar to the model of \citet{Sole2002} which is one-localized. We use independent $\text{Beta}\parenth{2,2}$ priors for the two parameters. The model reduces to probabilistic copying for $\theta_2=1$.

\subsection{\modelnumref.\ Jackson-Rogers model}

\citet{Jackson2007} proposed a growing model for social networks in which agents explore local neighborhoods to find friends in two stages. First, at each step, $m_\text{rnd}$ seeds $S_{t}$ are selected uniformly at random, and $\order_t$ is connected to each of them with probability $\theta_\text{rnd}$. Second, a set $C$ of $m_\text{nbr}$ nodes is chosen uniformly at random without replacement from the union of neighborhoods of seeds (even if $\order_t$ did not connect with them). Connections between each node in $C$ and $\order_t$ are created independently with probability $\theta_\text{nbr}$. The model is not localized because seed subgraphs are not necessarily connected to $\order_t$ if both $\theta_\text{rnd}$ and $\theta_\text{nbr}$ are small. We use independent flat $\text{Beta}\parenth{1,1}$ priors for $\theta_\text{rnd}$ and $\theta_\text{nbr}$. We set $m_\text{rnd}=m_\text{nbr}=10$ and initialize the growth process with a complete graph of $m_\text{rnd}+m_\text{nbr}+1$ as in \citet{Jackson2007}.

\subsection{\modelnumref.\ Watts-Strogatz model}

Similar to the connected small world model, the Watts-Strogatz model starts with a ring of nodes connected to their $z$ nearest neighbors~\citep{Watts1998}. Rather than adding edges, each edge in the ring is independently rewired with probability $\theta$. This model is not a growing network model and previously connected parts of the graph may become disconnected by rewiring if $\theta$ is large. We use a $\mathrm{Beta}\parenth{1,1}$ prior for the rewiring probability and set $z=4$.

\section{C.\ Tractability of connected small world for NDEs\label{app:connected-small-world-proof}}

The proposed NDE architecture can approximate the tractable posterior density of the connected small world model in \cref{eq:connected-small-world-posterior} well for non-trivial graphs, i.e., graphs that are neither empty nor complete.

\begin{enumerate}
    \item The first aggregation in \suppref{\cref{eq:gin-convolution}}{eq.~(5) of the main text} evaluates to $\mathbf{X}=\mathbf{\tilde{A}} \mathbf{1}=\mathbf{d}+1$, where $\mathbf{d}$ is the vector of degrees.
    \item Without loss of generality, we assume that each layer only has a single feature. Consider biases $b_1^{(1)}=0$, $b_2^{(1)}=-1$, and weights $W_1^{(1)}=\epsilon$, $W_2^{(1)}=1/\epsilon$. Then the MLP in \cref{eq:mlp} becomes
    \[
\phi^{(1)}\parenth{\mathbf{X}} = \tanh\parenth{ \epsilon\mathbf{X}}/\epsilon - 1.
    \]
    By Taylor expansion of $\tanh$ for small $\epsilon$, we get $\lim_{\epsilon\rightarrow0}\phi^{(1)}\parenth{\mathbf{X}}=\mathbf{X}-1=\mathbf{d}$.
    \item Setting $\gamma^{(1)} = 0$ in the residual connection, the first latent representation $\mathbf{H}^{(1)}$ is the vector of degrees.
    \item Due to residual connections in the network, we can propagate the degrees, up to a multiplicative constant, to the final GIN layer. The global mean-pooling layer evaluates the mean degree $\bar d$ which is a sufficient statistic that fully determines the posterior in \cref{eq:connected-small-world-posterior}.
    \item By the same argument, we can propagate the mean degree through dense layers with residual connections up to the final representation before the density estimation head.
    \item For sufficiently large values, the concentration parameters are approximately equal to the input of the softplus because $\mathrm{softplus}\parenth{x}\approx x$. Consequently, we can reconstruct the arguments of the beta distribution in \cref{eq:connected-small-world-posterior} well. In a sample of 100 simulated graphs, the smallest concentration parameter was $>40$, and the relative error of the linear softplus approximation is $<10^{-5}$ for $x\geq 10$.
\end{enumerate}

Consequently, any NDE in the family we consider can approximate the true posterior arbitrarily well, and any deficiency in performance is a result of improper convergence.